\documentclass{vgtc}                          

\ifpdf
  \pdfoutput=1\relax                   
  \usepackage{graphicx}                
  \DeclareGraphicsExtensions{.pdf,.png,.jpg,.jpeg} 
\else
  \usepackage{graphicx}                
  \DeclareGraphicsExtensions{.eps}     
\fi%

\graphicspath{{figures/}{pictures/}{images/}{./}} 

\usepackage{microtype}                 
\PassOptionsToPackage{warn}{textcomp}  
\usepackage{textcomp}                  
\usepackage{mathptmx}                  
\usepackage{times}                     
\usepackage{cite}                      
\usepackage{tabu}                      
\usepackage{booktabs}                  

\usepackage{balance}

\usepackage{amssymb}

\newcommand{\system}{\textit{VisKonnect}}

\usepackage{tikz}
\usepackage{xcolor}

\definecolor{myblue}{RGB}{47,85,151}
\newcommand*\rectBlue[1]{\tikz[baseline=(char.base)]{
            \node[shape=rectangle,fill=myblue, inner sep=2pt] (char) {\textcolor{white}{\small{#1}}};}}
            
\onlineid{0}

\vgtccategory{Research}

\vgtcinsertpkg

\title{Visually Connecting Historical Figures Through Event Knowledge Graphs}

\author{Shahid Latif
\and Shivam Agarwal
\and Simon Gottschalk
\and Carina Chrosch
\and Felix Feit
\and Johannes Jahn
\and Tobias Braun
\and Yanick Christian Tchenko
\and Elena Demidova
\and Fabian Beck\thanks{Latif, Agarwal, and Beck are with paluno, University of Duisburg-Essen (email: \{firstname.lastname\}@paluno.uni-due.de); Chrosch, Feit, Jahn, Braun, and Tchenko are students at University of Duisburg-Essen; Gottschalk is with L3S Research Center, Leibniz University of Hannover (email: gottschalk@L3S.de); Demidova is with Data Science \& Intelligent Systems (DSIS), University of Bonn (email: demidova@cs.uni-bonn.de). }} %

\teaser{
  \centering
  \includegraphics[width=\linewidth, trim= 0in 0.9in 0.in 0.2in]{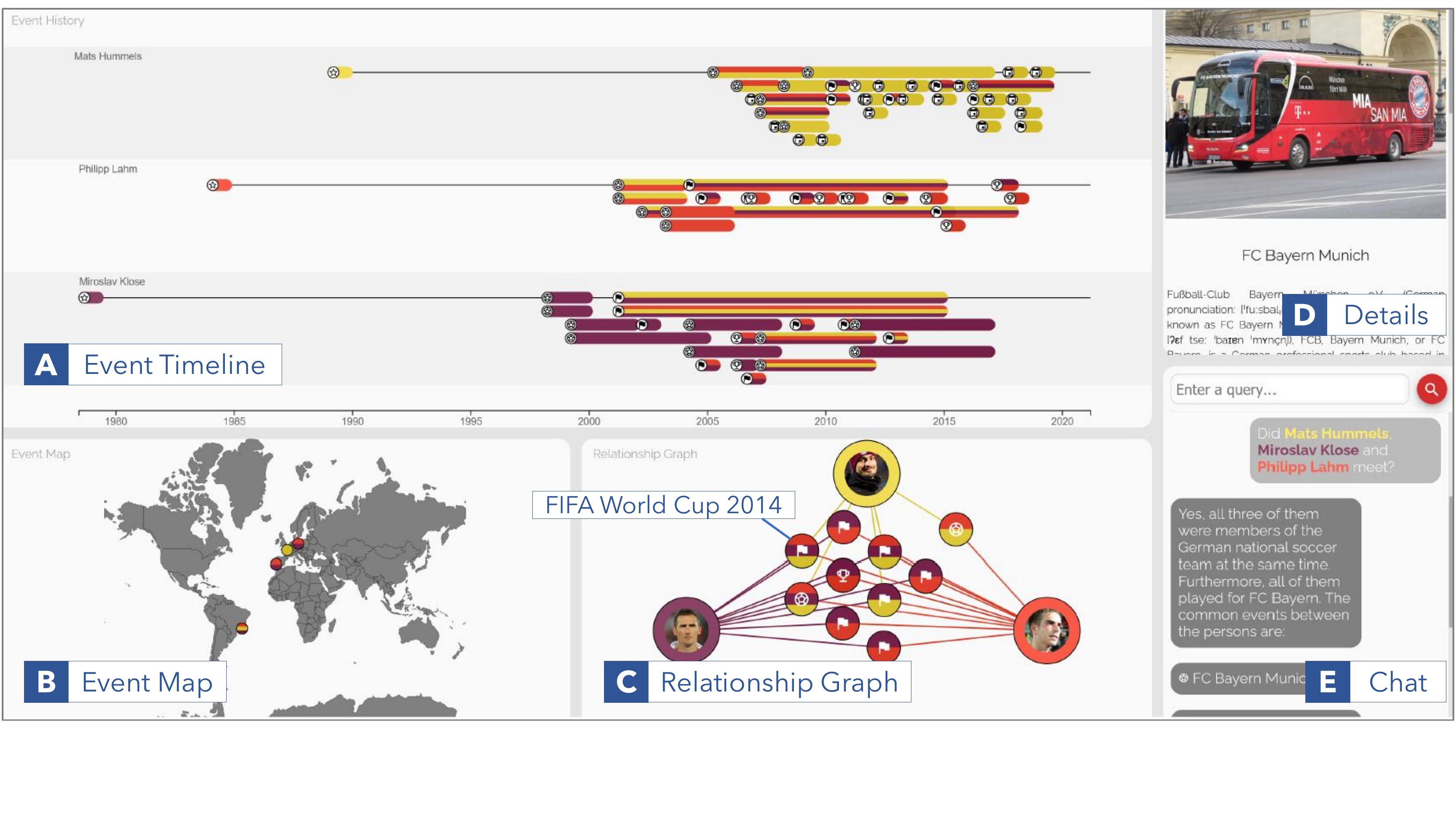}
  \caption{\system{} interface visualizing the result of a query on soccer players: (\textbf{A}) An event timeline shows individual (uni-colored rectangles) and shared (multi-colored rectangles) life events of all historical figures in a user query. (\textbf{B}) An event map provides a geographic perspective of those events. (\textbf{C}) A relationship graph provides an overview of all shared events. (\textbf{D}) Clicking an event brings up a related article from Wikipedia. (\textbf{E}) A chat interface enables typing in a query and generates a short textual answer.}
  \label{fig:teaser}
}

\abstract{
Knowledge graphs store information about historical figures and their relationships indirectly through shared events. We developed a visualization system, \system{}, for analyzing the intertwined lives of historical figures based on the events they participated in. A user's query is parsed for identifying named entities, and related data is retrieved from an event knowledge graph. While a short textual answer to the query is generated using the GPT-3 language model, various linked visualizations provide context, display additional information related to the query, and allow exploration.
} 

\begin{document}


\firstsection{Introduction}
\maketitle
History is significantly influenced, if not driven, by the interactions of a few outstanding individuals. Learning about their lives and how they interacted with each other can help us understand historic developments in world politics, sports, or science. Traditionally, information about connections between historic figures is reported in history books and encyclopedias, but limited to the connections drawn by the authors of the text. While text represents unstructured information, knowledge graphs store information about these persons in a structured way---for instance, their dates of birth, family ties, achievements, and participation in certain events---and can be used to infer possible connections between arbitrary figures of the same era. Search engines (e.g., Google) employ such knowledge graphs to show compact biographic information about historical figures as info boxes aside the search results. However, these simple excerpts from a knowledge graph are typically limited to a single person and do not convey interactions between different persons.

In this paper, we present \system{}, an approach that visualizes the connections between historical figures and their potentially intertwined lives. Such points of contact are events such as sport tournaments, award ceremonies, or summit meetings. \system{} relies on EventKG~\cite{gottschalk2019eventkg}, an event-centric knowledge graph that covers relations between events and persons and allows inferring the required information. Through three visualizations---namely, an event timeline, an event map, and a relationship graph---\system{} provides access to these connections. Addressing a wider audience, we use a chat interface to query the data naturally and to provide a direct answer to the user's information need. \autoref{fig:teaser} shows the response of an example query asking whether three soccer players (Mats Hummels, Miroslav Klose, and Philipp Lahm) have met. While the chat explicitly answers the question (\autoref{fig:teaser} \rectBlue{E}), three-colored rectangles in the timeline visualization (\autoref{fig:teaser}~\rectBlue{A}) reveal the shared events. 

\section{Related Work}

Our work is connected to visualization of knowledge graphs, visualization of biographies and people's interactions, as well as natural-language interfaces for visualization systems. 

Cross-domain knowledge graphs such as Wikidata and DBpedia typically convey world knowledge at a large scale. The objects in a knowledge graph are typically explored using graph-based, text-based, or mixed interfaces~\cite{desimoni2020empirical, antoniazzi2018rdf}. In case of existence of specific properties such as coordinates, more specialized tools with a spatio-temporal focus can be employed~\cite{ikkala2021sampo}. In \system{}, we are bringing these different perspectives together for a special use case (namely, a temporal, a map-based, and a graph-based visualization, as well as a text-based interface).

Joseph Priestley’s work on ``Chart of Biography''\cite{priestley1765chart} includes a hand-sketched visualization of the lifespans of prominent people using a horizontal timeline. Each person is represented in a specific row as a horizontal line and grouped in one of the six categories based on their profession. Since then, several techniques have been proposed to visually explore the biographies of historical figures. For instance, based on Priestley’s design, Khulusi et al.~\cite{khulusi2019interactive} proposed interactive biographical charts for detailed exploration. Leskinen et al.~\cite{leskinen2018analyzing} built visualizations enabling faceted search of the biographical data. Some recent works (e.g.,~\cite{gottschalk2020eventkg+,hyvonen2014life, han2021hisva}) visualize biographical data enriched with events in the lifespans of historical figures. However, unlike our approach, these techniques do not explicitly represent the shared events connecting the lives of historical figures in the proposed visualizations. 

Recent works have proposed systems that integrate template-based natural-language text with visualizations to both explain and explore structured data (e.g.,~\cite{mumtaz2020exploranative,latif2019vis}). It is also possible to inspect structured data using natural-language queries~\cite{bacci2020inspecting,bowen2020flowsense, kumar2016towards,sun2010articulate}. These systems focus on understanding the natural-language queries and primarily output standard visualization (e.g., bar and line chart). However, when a query is complex and consists of multiple entities, it becomes challenging to retrieve and visualize the relevant data. In our approach, we address the challenge by using a chat interface to provide direct response to the natural-language queries, in addition to the linked customized visualizations for context and details.

\begin{figure}
  \centering
  \includegraphics[width=\linewidth]{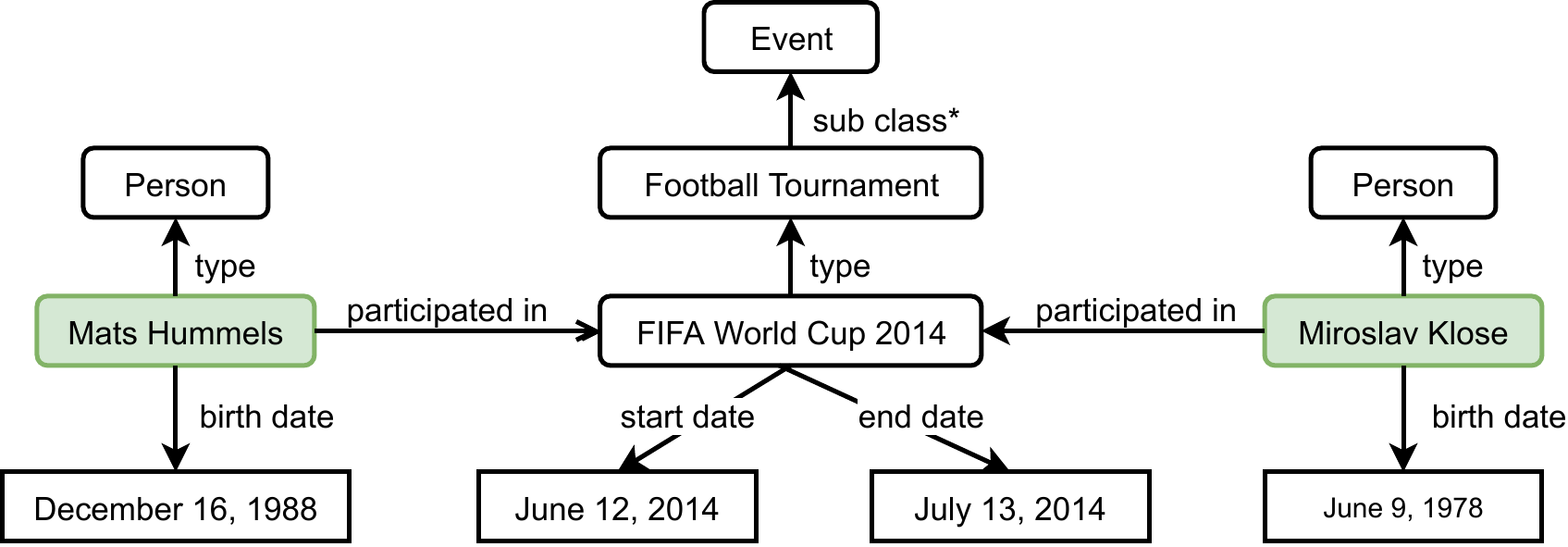}
  \caption{An excerpt of EventKG about the relationship between Mats Hummels and Miroslav Klose; \textit{sub class}$^{\ast}$ denotes transitivity.}
  \label{fig:kg_example}
\end{figure}

\section{\system{}: The System}
Given a natural-language question about two or more persons of public interest, \system{} returns a textual answer to the question plus visualizations that reveal potential connections between the persons mentioned in the question. As shown in \autoref{fig:teaser}, these visualizations focus on different facets: \rectBlue{A} the event timeline represents the persons' lives over time, \rectBlue{B} the map view depicts geographic co-locations, and \rectBlue{C} the relationship graph provides a concise overview of shared events. Additionally, \rectBlue{E} the chat panel serves as a textual question--answer interface. \system{} is a web-based system; the front end is developed in TypeScript using Angular while the backend is written in Python. 

We followed a design process where researchers from two fields (three from visualization and two from knowledge graph research) supervised and contributed to a student-led project. The overarching vision and goal of the project was jointly developed by the researchers. Five graduate students (all co-authors) were mainly responsible for the implementation of the prototype. After being introduced to the topic, they initially interviewed one visualization expert and one knowledge graph expert (both authors of this paper) to derive the requirements. Based on these interviews, mock-ups were designed and discussed. 

\subsection{EventKG and Data Retrieval}

As an underlying resource for obtaining data about the lives of historical figures, \system{} requires structured information about real-world entities, specifically persons, their relationships, and the events they have been involved in. A natural source of this information are knowledge graphs. A knowledge graph $G=(N,R)$ consists of a set of nodes ($N$) representing real-world entities (e.g., \textit{Mats Hummels}, the \textit{FIFA World Cup 2014}) and a set of edges ($R$) denoting relationships between entities~\cite{hogan2020knowledge}. These relationships are represented as triples consisting of a subject, a predicate, and an object (e.g., $\langle$Mats Hummels, participated in, FIFA World Cup 2014$\rangle$ and $\langle$Mats Hummels, type, Person$\rangle$). 

\system{} utilizes EventKG~\cite{gottschalk2019eventkg}, a knowledge graph incorporating event-centric and temporal information extracted from several other knowledge graphs and semi-structured sources such as Wikipedia. EventKG contains several types of events, temporal relationships, and an event class ontology that builds a basis for the visualization components of \system{}. 
\autoref{fig:kg_example} shows an excerpt of EventKG with triples about the soccer players Mats Hummels and Miroslav Klose that indicate a shared event: Even though there is no direct relationship between the players, we can still infer that they both participated in the FIFA World Cup 2014.

\begin{figure}[t]
  \centering
  \includegraphics[width=1.0\columnwidth, trim= 0.1in 1.3in 3.8in 0in]{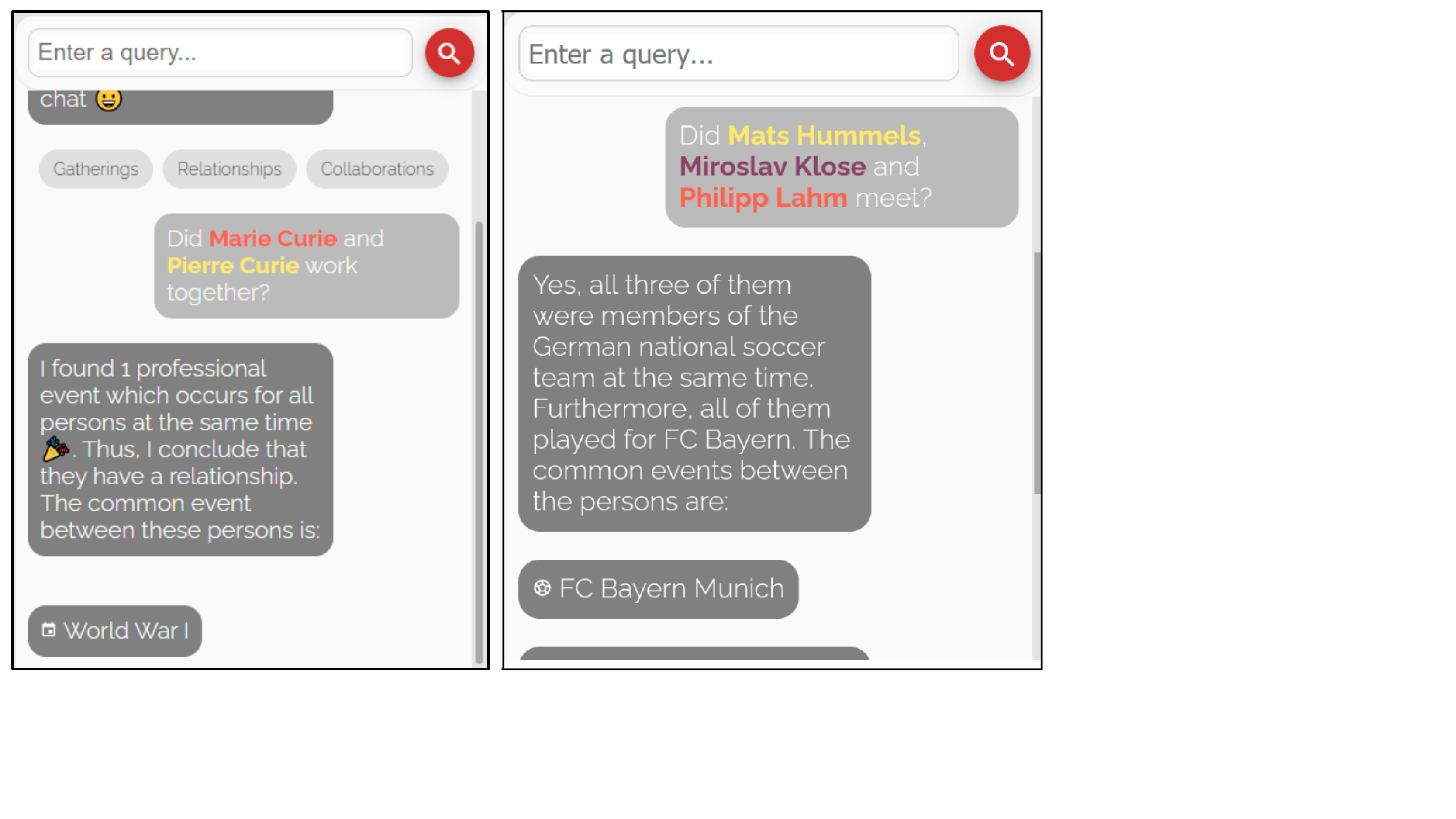}
  \caption{Chat responses based on templates (left) and GPT-3 (right).}
  \label{fig:chat}
\end{figure}

Retrieving information from knowledge graphs is challenging for many potential users as it requires both knowledge of the SPARQL query language~\cite{sparql} and the specific schema of the given knowledge graph. Hence, a major goal of \system{} is to relieve the user of formal query formulation and instead provide a natural-language interface. Upon a user's query about persons of interest, \system{} retrieves the following information from the EventKG: (i) The relationships between a person and related events (e.g., Mats Hummels' participation in the FIFA World Cup 2014). For such events, we retrieve their labels, types, start and end dates, as well as their coordinates, whenever possible. If multiple persons in the query are related to the same event, we consider it as a \textit{shared event}. (ii) The temporal relationships between a person and another person (e.g., Marie Curie was married to Pierre Curie from 1895 to 1906). 

To translate a natural-language user query to SPARQL, we extract the names of persons and try to detect the query intent. We use parts-of-speech tagging and named entity recognition using pre-trained models of \textit{spaCy} to identify person names. These names are then matched against EventKG through ElasticSearch~\cite{elasticsearch}, and a SPARQL query is formulated. The constructed SPARQL queries are generic---independent of user intent---and retrieve all individual and shared events for the queried persons.

The intent of the question is needed to generate a textual answer. In our context, \textit{intent} relates to the type of relationships between persons (and between persons and events) users are interested in. We restrict ourselves to three types of relationships between persons and events: professional, personal, and general relations. Professional relations describe working alliances between people, such as receiving an award or war alliances. Personal relations consist of friendships between persons or family ties identified by events like marriage or childbirth. Any other relation falls under the category of general relation (e.g., if a person has influenced the other person). We use a rule-based approach to detect the intent. These rules are identified through the analysis of a set of 500 questions that are relevant for EventKG~\cite{souza2020event}; each relationship type is associated with a set of pre-defined words. For instance, words like ``collaborate'', ``work'', ``ally'' define professional relations. To classify the user's query as one of the relationship type, we use semantic similarity between characteristic words of a relationship type and important words in the user query (excluding nouns and stop words). We use a pre-trained \textit{word2vec} model~\cite{NIPS2013_9aa42b31} for computing semantic similarity. The intent is only relevant for generating template-based answers and has no impact on visualizations.


\begin{figure*}[t]
  \centering
  \includegraphics[width=\linewidth, trim= 0.2in 4.7in 0.2in 0in, clip=true]{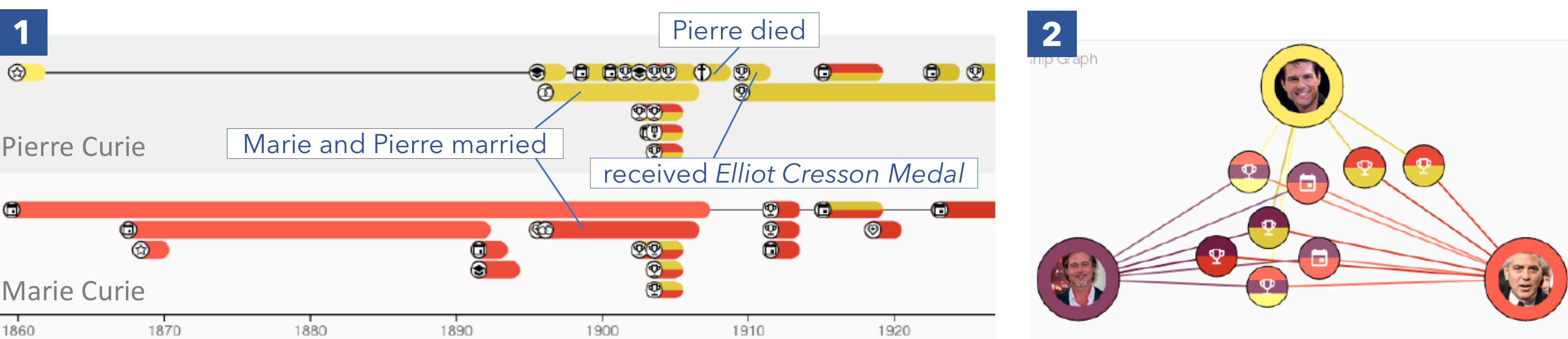}
  \caption{Cut-outs of two responses generated by \system{}. (\textbf{1}) Event timeline for Example 2. (\textbf{2}) Relationship graph for Example 3.}
  \label{fig:examples}
\end{figure*}

\subsection{Interface}
\system{}'s interface consists of the following components (\autoref{fig:teaser}):

\textbf{Event Timeline.} The event timeline (\autoref{fig:teaser}~\rectBlue{A}) provides an overview of the overlapping events for all persons in a user query. We use a distinct color for each person and vertically separate the persons into individual timelines. Each timeline begins with the person’s date of birth and ends at either the date of death or the last relevant event (in case the person is still alive or received awards posthumously). Events are represented by thick colored lines, with their duration encoded in the length. We use exoteric icons and place them at the start of the line to communicate the type of events, for instance, a star icon for birth, a cross icon for death, a flag icon for sports tournaments. Many events can take place at the same time in a person's lifetime, therefore the overlapping events for the same person are placed on separate stacked rows. The shared events are visualized as multi-colored lines, with colors corresponding to the persons involved.

\textbf{Event Map.} The event map (\autoref{fig:teaser}~\rectBlue{B}) gives a geographical perspective of places where the events took place. For instance, \autoref{fig:teaser}~\rectBlue{B} shows that all three soccer players participated in an event in Brazil (FIFA World Cup 2014). The events are displayed as colored circles, which are horizontally divided in different colors if the events are shared among two or more people. Since not all events have a geographical location in EventKG, fewer events might appear on the map compared to other views.

\begin{figure}
  \centering
  \includegraphics[width=\columnwidth, trim= 0in 4.8in 6.75in 0.1in, clip=true]{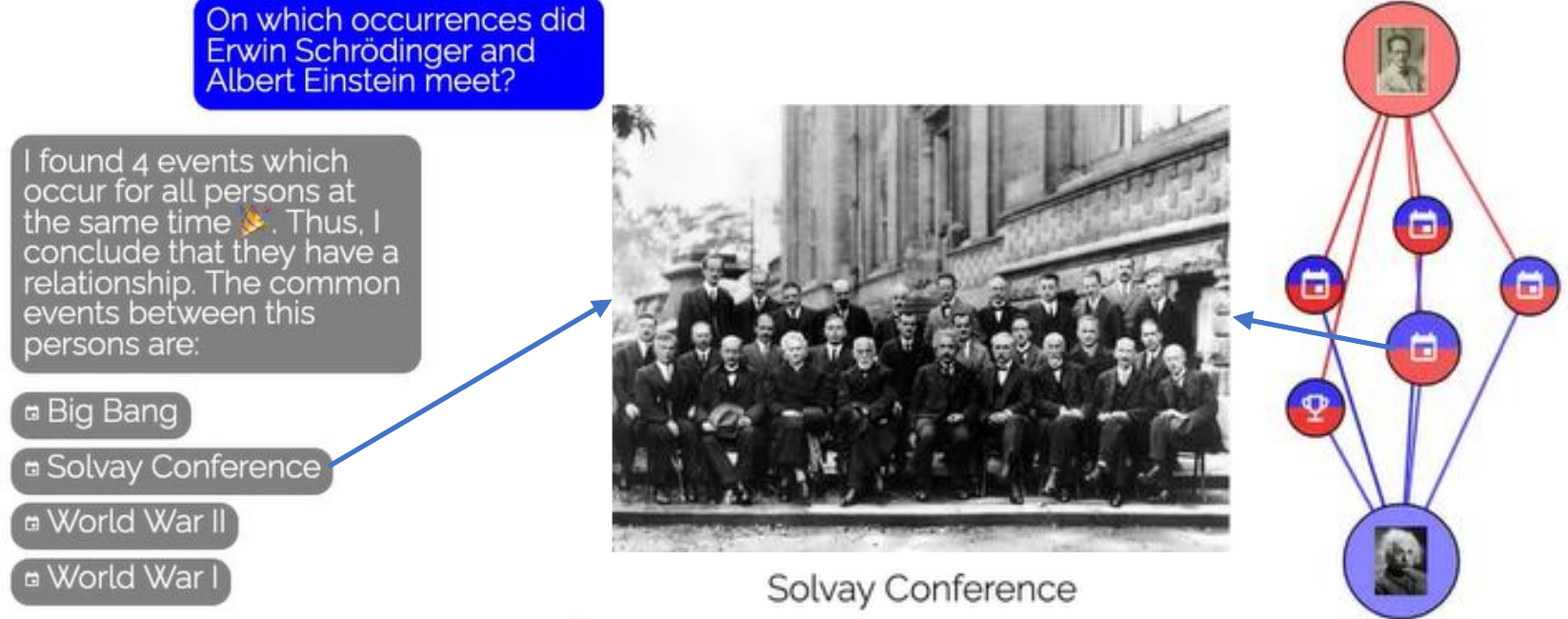}
  \caption{Excerpt of a question about Einstein and Schrödinger (Example 4). The template-based answer identifies Solvay conference as a shared event where the two scientists were photographed together.}
  \label{fig:example3}
\end{figure}

\textbf{Relationship Graph.} The relationship graph (\autoref{fig:teaser}~\rectBlue{C}) shows the connectivity of queried persons through shared events as a force-directed graph. The persons are displayed with their picture and corresponding background color as assigned in the event timeline (\autoref{fig:teaser}~\rectBlue{A}). 
The events are connected to corresponding persons via colored edges. Nodes representing shared events use multiple colors similar as in the event map; events only assigned to a single person are not shown to reduce clutter. A desired consequence of the force-directed layout is that the more connected event nodes appear towards the center, while the person nodes float in the periphery.

\textbf{Chat.} The chat (\autoref{fig:teaser}~\rectBlue{E}) serves as the entry point to the system. It greets users with a welcome message and a few sample queries. Users can freely type in a query. \autoref{fig:teaser}~\rectBlue{E} shows an example query and the generated response. Answering the user’s question is done either using the GPT-3---an autoregressive language model that uses deep learning to produce human-like text~\cite{DBLP:conf/nips/BrownMRSKDNSSAA20}---or through the identified shared events with pre-written text templates. GPT-3 receives the user’s question as input and tries to generate an answer based on its 499 billion tokens of pre-trained data. Since these tokens come from a variety of sources, such as Wikipedia or Common Crawl, GPT-3 answers are often more precise than the information displayed by the template. If GPT-3 fails to answer the question, \system{} generates a response from the text templates. This response is based on the number of identified events and the intent. When there is no identified event, \system{} reports that no event was found with temporal overlap between the queried persons. When shared events are detected, the answer reflects this and the identified events are displayed below. \autoref{fig:chat} shows two textual answers generated by GPT-3 and the text templates, respectively. 

\system{} uses consistent colors and icons across all views to support quickly moving from one view to another. Furthermore, interactions link the views. For instance, hovering over an event in any of the view highlights related events in all other views. Users can click any event anywhere to pull up the related Wikipedia article in a dedicated panel (\autoref{fig:teaser}~\rectBlue{D}). Besides, zooming and panning in the event timeline, event map, and relationship graph helps in exploring the details.

\section{Examples}

To evaluate the expressiveness and usefulness of \system{}, we tested the system by asking various queries. Here, we first present a few examples and then analyze the system’s response to highlight benefits and potential problems.

\textbf{1. Did Mat Hummels, Miroslav Klose, and Philipp Lahm meet?} \system{} (GPT-3) states that they all were members of the German national soccer team at the same time, and they all played for FC Bayern. The chat lists two shared events: \textit{``FIFA World Cup 2014''} and \textit{``FC Bayern''} (\autoref{fig:teaser}~\rectBlue{E}). This fact also becomes evident by looking at two three-colored events in the relationship graph \rectBlue{C}. The event timeline \rectBlue{A} further reveals that the three soccer players participated in other sports events together between 2002 and 2020. 

\textbf{2. When did Pierre Curie and Marie Curie marry?} The system (GPT-3) correctly answers the question as: ``They married in July 1895'' followed by listing two individual events \textit{``spouse: Marie Curie''}, and \textit{``spouse: Pierre Curie''}. Clicking the ring icon in their lifelines, it is found that the marriage lasted from 1895 to 1906; Pierre died in 1906 (\autoref{fig:examples}~\rectBlue{1}). Their shared events include \textit{``Nobel Prize in Physics (1903)''}, \textit{``Davy Medal (1903)''}, and \textit{``World War I''}. Pierre is also connected to events after his death. For instance, he posthumously received the ``Elliot Cresson Medal (1909)''. 

\textbf{3. Did Brad Pitt, George Clooney, and Tom Cruise all receive movie awards?} The system (GPT-3) answers that they all received the Golden Globe Award and were all nominated for an Oscar award. Two events appear in the chat: The \textit{``Golden Globe Award''} and \textit{``Academy Awards''} (Oscar) supporting the claim generated by GPT-3. Searching for the award symbol in the other views (e.g., relationship graph -- \autoref{fig:examples}~\rectBlue{2}) and looking up details further asserts the answer.

\textbf{4. Did Erwin Schrödinger meet Albert Einstein?} \autoref{fig:example3} shows a cut-out of the generated response. The template-based answer describes that four shared events were identified for the two scientists; they can also be seen in the relationship graph on the right. Among these, the \textit{``Solvay Conference''} seems the most likely event for a true meeting---clicking it actually brings up a photograph where both men can be seen together.

\system{} could answer most questions correctly or partially correctly as text; the answers generated by GPT-3 were more precise than the template-based answers. The accompanying explorable visualizations helped verify the answer, provided context, and assisted in navigating to the answer in case \system{} failed to provide a textual answer. On the downside, we observed that, sometimes, GPT-3 answers did not fully align with the visualized events. Sometimes, events are very general and overestimate who might have met or is related (e.g., \textit{World War I} is assigned to many persons). As some events are prolonged or assignment to an event does not necessarily imply the physical presence of the person. Some events may seem misplaced in the event timeline and start either before birth (e.g., before Marie Curies's birth in \autoref{fig:examples}~\rectBlue{1}) or after death (e.g., after Pierre Curie's death in the same figure). \system{} also had problems recognizing names of some historical people, especially the ones having titles, prefixes, or suffixes (e.g., Queen Elizabeth II).

\section{Discussion}

We consider our approach as a proof of concept that a natural-language interface integrated with knowledge graphs and visualizations can support lay users in answering non-trivial questions about historical figures. Still, there are various open issues and interesting research questions to be addressed.

\textbf{Verification and Explanation of Results.} Knowledge graphs and language models are two distinct approaches and have recently been combined for answer generation~\cite{kalo2020knowlybert}. \system{} demonstrates how this combination---through linked visualizations and textual answers---can provide immediate benefit to the user. With \system{}, on the one hand, a user can verify the answer generated by the language model through visualizations of related events. On the other hand, the generated answer helps in explaining the information conveyed through the visualization of knowledge graph data.

\textbf{Exploration of Intertwined Lives.} \system{} showcases the biographies of queried people. The linked visualizations enable exploration of their lives not only from temporal and spatial perspectives, but also highlight the shared events to show how their lives were intertwined. The consistent color encoding across the entire interface allows easy navigation. This ability to explore relationships in the lives of historical figures can be applied in education, as demonstrated by the usage scenarios of the \textit{HisVA} system \cite{han2021hisva}.

\textbf{Relevant Information Selection.} Knowledge graphs typically contain millions of nodes and edges (for instance, EventKG has more than two million relations connecting persons to events). Consequently, the selection of relevant pieces of information that are shown to the user poses an important challenge for any visualization built on top of knowledge graphs. For the selection of relevant events, \system{} relies on the number of mentions of the respective events on Wikipedia. As a result, events that are of large historic significance such as the First or Second World War play an important role in \system{}, independent of the context, i.e., the question and the persons involved. For a more focused exploration of the persons' lives, there is a need for context-dependent relevancy metrics and for providing interaction possibilities that allow users deselecting specific or less relevant pieces of information. Another problem relates to the ambiguity in entity names (e.g., whether the user is interested in Winston Churchill, the British prime minister or the American novelist). Having human interventions in case of such ambiguities through interactive widgets (similar to Eviza~\cite{setlur2016eviza}) may resolve the problem to some extent.

\textbf{Visual Scalability.} 
Taking inspiration from \textit{TimeSets}~\cite{nguyen2016timesets}, we used unique colors for each queried person and represented the shared events by filling the respective colors in the glyphs (e.g., lines in the timeline and nodes in the relationship graph). As a result, it enables the users to infer the shared events among queried historical figures. However, since it may become difficult to differentiate between more than five colors, the visual scalability of the interface is limited to about five people in a single query. Besides, the event timeline can quickly become visually cluttered when a person is linked to many events that temporally overlap.



\textbf{Answers by Text Templates or GPT-3?} While the use of GPT-3 in the chat allows precise answers (see \autoref{fig:chat}), it is hard to control, predict what it would generate, and therefore integrate with the other visual views, which use EventKG as underlying data. Potential differences in textual and visual answers might confuse the users. In contrast, answers generated by text templates can always be made consistent to the visual answer. However, it requires bigger effort to make them sound natural and handle all possible cases, given the user has all the freedom to ask questions. While \system{} leaves it up to the user to verify whether GPT-3 produced the right answer, a future step is to automatically analyse GPT-3 response to flag potential misinformation or conflicting information.



\section{Conclusion}
We presented an approach to visualize and explore intertwined lives of historical figures using event knowledge graphs. The \system{} system provides an accessible way of querying the data through natural language and visualizes the events that connect persons. While the chat explicitly answers the question, the visualizations help in verifying the answer and provide context. 

\acknowledgments{
This work is partially funded by the German Research Foundation (DFG, Deutsche Forschungsgemeinschaft) under the projects ``vgiReports'' (424960846) and ``WorldKG'' (424985896), and the Federal Ministry of Education and Research (BMBF), Germany under ``Simple-ML'' (01IS18054).
}

\balance
\bibliographystyle{abbrv-doi-hyperref}

\bibliography{references}
\end{document}